\documentclass[journal]{IEEEtran}
\usepackage[utf8]{inputenc} 
\usepackage{hyperref}       
\usepackage{url}            
\usepackage{booktabs}       
\usepackage{amsfonts}       
\usepackage{nicefrac}       
\usepackage{microtype}      
\usepackage{graphicx}
\usepackage{enumerate}
\usepackage[dvipsnames]{xcolor}
\usepackage{subfig}
\usepackage{titlesec}
\usepackage[noadjust]{cite}
\usepackage[normalem]{ulem}
\usepackage[colorinlistoftodos]{todonotes}
\usepackage{longtable}

\ifCLASSINFOpdf
\else
\fi
\usepackage{amsmath}
\begin{document}
%
\title{Carbon-Aware Computing for Datacenters}
%
%
%

\author{Ana Radovanovi\'c, 
        Ross Koningstein, 
        Ian Schneider,
        Bokan Chen,
        Alexandre Duarte,
        Binz Roy,
        Diyue Xiao,
        Maya Haridasan,
        Patrick Hung,
        Nick Care,
        Saurav Talukdar, 
        Eric Mullen, 
        Kendal Smith,
        MariEllen Cottman,
        and Walfredo Cirne
\thanks{The authors are with Google, Inc. Mountain View, CA, 94043 (Email: anaradovanovic@google.com, ross@google.com, ischneid@google.com, bokanchen@google.com, alexandredu@google.com, binzroy@google.com, diyuexiao@google.com, haridasan@google.com, phfhung@google.com, ncare@google.com, stalukdar@google.com, ericmullen@google.com, kendalsmith@google.com, meacottman@google.com, walfredo@google.com)
.}
}

\maketitle

\begin{abstract}
The amount of CO$_2$ emitted per kilowatt-hour on an electricity grid varies by time of day and substantially varies by location due to the types of generation. Networked collections of warehouse scale computers, sometimes called Hyperscale Computing, emit more carbon than needed if operated without regard to these variations in carbon intensity. This paper introduces Google's system for Carbon-Intelligent Compute Management, which actively minimizes electricity-based carbon footprint and power infrastructure costs by delaying temporally flexible workloads. The core component of the system is a suite of analytical pipelines used to gather the next day's carbon intensity forecasts, train day-ahead demand prediction models, and use risk-aware optimization to generate the next day's carbon-aware Virtual Capacity Curves (VCCs) for all datacenter clusters across Google's fleet. VCCs impose hourly limits on resources available to temporally flexible workloads while preserving overall daily capacity, enabling all such workloads to complete within a day. Data from operation shows that VCCs effectively limit hourly capacity when the grid's energy supply mix is carbon intensive and delay the execution of temporally flexible workloads to “greener” times.
 \end{abstract}


%
\IEEEpeerreviewmaketitle


\section{Introduction}\label{sec:intro}
Demand for computing resources and datacenter power worldwide has been continuously growing, now accounting for approximately 1\% of total electricity usage \cite{masanet2020recalibrating}. Between 2010 and 2018, global datacenter workloads and compute instances increased more than sixfold \cite{masanet2020recalibrating}. In response, new methodologies for increasing datacenter power and energy efficiency are required to limit their growing environmental, economic and performance impacts \cite{dayarathna2015data, urs2020efficiency}.
 
The datacenter industry has the potential to facilitate carbon emissions reductions in electricity grids. A considerable fraction of compute workloads have flexibility in both when and where they run. Given that emissions from electricity production vary substantially by time and location \cite{li2017marginal, thind2017marginal, callaway2018location, Khan2019temporal}, we can exploit load flexibility to consume power where and when the grid is less carbon intensive. By effectively managing its load, the datacenter industry can contribute to a more robust, resilient, and cost-efficient energy system, facilitating grid decarbonization. Electric grid operators, in turn, can possibly benefit by as much as EUR 1B/year \cite{EPEXReport2020}.

Furthermore, shifting execution of flexible workloads in time and space can decrease peak demand for resources and power. Since datacenters are planned based on peak power and resource usage, smaller peaks reduce the need for more capacity. Not only does this save money, it also reduces environmental impacts.

This paper describes the methodology and principles behind Google's system for Carbon-Intelligent Compute management, which reduces grid carbon emissions from Google's datacenter electricity use and reduces operating costs by increasing resource and power efficiency. To accomplish this goal, the system harnesses the temporal flexibility of a significant fraction of Google's internal workloads that tolerate delays as long as their work gets completed within 24 hours. Typical examples of such workloads are data compaction, machine learning, simulation, and data processing (e.g., video processing) pipelines -- many of the tasks that make information found through Google products more accessible and useful. Note that other loads include user-facing services (Search, Maps and YouTube) that people rely on around the clock, and our cloud customers' workloads running in allocated Virtual Machines (VMs), which are not temporally flexible and therefore not affected by the new system. 

Workloads are comprised of compute jobs. The system needs to consider compute jobs' arrival patterns, resource usage, dependencies and placement consequences, which generally have high uncertainty and are hard to predict (i.e., we do not know in advance what jobs will run over the course of the next day). Fortunately, in spite of high uncertainties at the job level, Google's flexible resource usage and daily consumption at a cluster-level and beyond have demonstrated to be quite predictable within a day-ahead forecasting horizon. The aggregate outcome of job scheduling ultimately affects global costs, carbon footprint, and future resource utilization. The workload scheduler implementation must be simple (i.e., with as little as possible computational complexity in making placement decisions) to cope with the high volume of job requests. 

The core of the carbon-aware load shaping mechanism is a set of cluster-level \cite{43438} Virtual Capacity Curves (VCCs), which are hourly resource usage limits that serve to shape each cluster resource and power usage profile over the following day. These limits are computed using an optimization process that takes account of aggregate flexible and inflexible demand predictions and their uncertainty, hourly carbon intensity forecasts \cite{tomorrowwebsite}, explicit characterization of business and environmental targets, infrastructure and workload performance expectations, and usage limits set by energy providers for different datacenters across Google's fleet. 

The cluster-level VCCs are pushed to all of Google's datacenter clusters prior to the start of the next day, where they set hourly limits for total flexible compute usage. These limits directly impact the real-time admission of flexible workloads. As a consequence, at times of day when the local grid's carbon intensity is expected to be high, those clusters VCC values tend to be smaller, which reduces their total compute and power usage (for an example and discussion on the load shaping mechanism, see Subsection \ref{sbsec:mechanism}). The reduction of usage is achieved via delaying the execution of flexible computing tasks to later times of the day.

\begin{figure}
  \centering
  \includegraphics[width=\linewidth]{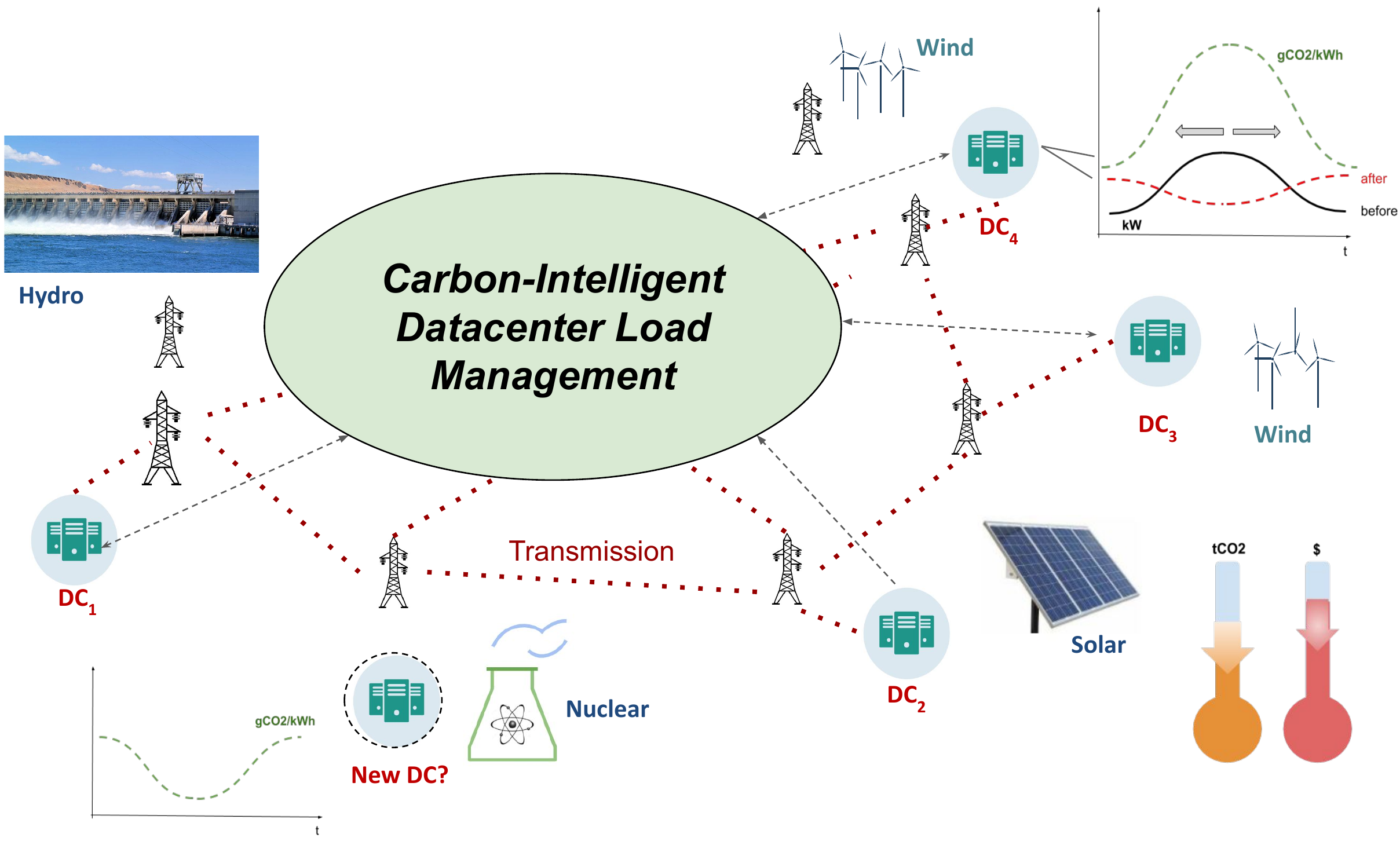}
  \caption{Carbon-Intelligent management of Google's datacenter portfolio. The green dashed curves associated with specific datacenter locations are intraday carbon intensities of the local electricity generation. The impact of CICS on intraday datacenter power consumption is depicted using black (original) and dashed red (shaped) curves. With CICS, datacenter power consumption during peak carbon intensity hours is intended to be lower than without CICS.}
  \label{overview}
\end{figure}

The scope of the Carbon-Intelligent Computing System (CICS) presented in this paper is global; the guidelines (VCC curves) shift load to lower overall carbon impact wherever Google locates its data centers, regardless of the generation source of local lower-carbon energy (Figure \ref{overview}). This helps Google achieve its environmental, efficiency, and performance targets across the world.

While datacenter power load shaping has been the subject of theoretical treatments and smaller scale prototypes in the past \cite{5598305, 6877627, Bianchini2014GreenComputing, zhenhua2011Sigmetrics, zhenhua2015IEEENetw, zhenhua2012Sigmetrics, 6114408, GreenHadoop}, the CICS is novel in the way it combines the following properties: 
\begin{itemize}
\item \textit{First-of-its-kind demonstration}: To the best of our knowledge, this is the first demonstration of carbon-aware algorithms that shifts datacenter computing in time to realize global environmental and efficiency objectives \cite{radovanovic2020our}, using automated adjustments based on current and forecasted grid conditions.
\item \textit{Effective, risk-aware, optimization approach}: Our load shaping mechanism uses a risk-aware optimization approach that incorporates fleetwide, cluster-level load forecasts, power models \cite{radovanovic2021power}, explicit workload and infrastructure performance expectations, as well as datacenter power-contract information. The demonstrated effectiveness of the proposed mechanism is due to the high accuracy of the cluster-level load forecasts and power models. 
\item \textit{Generic compute usage characterization}: The optimization model has a more generic notion of uncertainty than existing approaches, to the best of our knowledge. There is no a-priori defined model for job arrival/request rates.
\item \textit{Extensible analytical treatment}: The approach is extensible since its optimization formulation can be modified to incorporate different business objectives and operational constraints, including characterizations of spatially flexible usage. Furthermore, day-ahead management of load usage limits can be potentially leveraged to dynamically turn machines on/off to further increase power efficiency.  
\item \textit{Reliable}: The system is built with reliability principles in mind, with an established monitoring and feedback loop which ensures that the system operates as intended and that it adapts to changes in resource usage trends and system upgrades. Flexible load management is subject to explicitly defined Service Level Objectives (SLOs), and it does not impact the performance of the inflexible workload.
\item \textit{Modular}: Due to the nature of Google's workload, the day-ahead, load-shaping optimization runs independently from real-time job-level scheduling, long-term resource planning, and user-level interfaces with the compute infrastructure. 
\item \textit{Scheduler-agnostic}: Since CICS provides only the constraints used by real-time job scheduling, the real-time job scheduler can evolve independently of the system.
\end{itemize}

The paper is organized as follows. After the introduction and the overview of the related research in Section \ref{sec:intro}, Section \ref{sec:overview} provides basic information on Google's datacenter power architecture, as well as the key concepts behind the real-time management of compute workload. The end of Section \ref{sec:overview} provides a high level introduction and the key design principles behind the mechanism implemented to shape Google's datacenter compute load and the related power usage in a carbon- and efficiency-aware manner. Section \ref{sec:shapinganalytics} contains the details of the analytic pipelines that enable the load shaping mechanism. Section \ref{sec:demo} discusses how clusters' compute utilization, amount of flexible load, prediction uncertainty of flexible/inflexible demand, and intraday variation in carbon intensity impact the effectiveness of this carbon-aware load shaping mechanism. The paper is concluded in Section \ref{sec:conclusion}.

\subsection{Related research}
Greenhouse gas emissions from electricity production vary substantially by time and location \cite{li2017marginal, thind2017marginal, callaway2018location, Khan2019temporal}. This wide variation in carbon intensity (average greenhouse gas emissions per unit of energy consumption) or marginal emissions (the additional greenhouse gas emissions per additional unit of electricity consumption) imply that the time and location of electricity consumption has a large effect on its associated global warming impact. This variation has implications for electric vehicle charging \cite{zivin2014spatial}, energy efficiency \cite{boomhower2020energy} and datacenter energy consumption. Assessing and forecasting system conditions to improve the scheduling of flexible demand have been identified as high-priority areas for machine learning efforts to combat climate change \cite{rolnick2019tackling}. 

Due to the high variability of electricity-related carbon dioxide emissions, temporal or spatial flexibility of electricity demand can be harnessed to help reduce grid emissions. Theoretical work focuses on the control and optimization of flexible consumption; the objective can focus on carbon, electricity costs, or grid stability (or some combination, as in \cite{olsen2019profitable}) without significant difference in the theoretical treatment. Example of theoretical frameworks for flexible load include control methods \cite{GELAZANSKAS201422} using linear time-invariant control systems \cite{7063922}, a Markov decision process with deferrable loads \cite{6760990}, decentralized control \cite{soton271985}, and distributed dynamic online convex optimization \cite{lesage2020dynamic}. These methodologies are applied to a variety of load types, including electric vehicle charging \cite{6081962}, grid-connected battery storage \cite{salas2018benchmarking, 8340703}, and thermostats \cite{6669582, 6832599, 8618975}. In our implementation, we use a centralized optimization-based approach to control datacenter loads, with an objective to reduce carbon emissions and increase resource efficiency.

Carbon-aware datacenter load management is not a novel concept. In addition to a wide range of work focusing on reducing energy consumption of datacenter hardware equipment, including IT, cooling, and supply systems (see \cite{urs2020efficiency}, and information on the standardized actuators for energy-driven management of IT equipment in \cite{acpi}), it has been recognized that sustainable datacenters require intelligent and unifying solutions for energy-aware management of both datacenter hardware and software architectures \cite{Jordi2017}. It was also acknowledged that energy-aware datacenters need to enable (i) awareness of carbon footprint, (ii) efficiency of IT, cooling and power supply altogether, (iii) increased use of renewable energy, (iv) increased energy market participation, and (v) energy strategy that is aligned to datacenter load's flexibility \cite{Jordi2017}.

There have been numerous theoretical treatments, small-scale prototypes and simulation-based studies proposing designs for datacenter compute load management with the goal to minimize datacenter portfolio energy cost and environmental objectives. The majority of the research has focused on self-managed datacenters, i.e. where a single company manages all the infrastructure necessary to support their computing needs. There are also some recent proposals of stylized, decentralized optimization models to incentivize colocation tenants to carbon- and cost-effectively manage their power demand \cite{7573118,7039140}.  

Carbon- and cost-aware compute management can be achieved either via shifting workloads across datacenter locations, or by delaying jobs' execution. A body of research has focused on real-time rebalancing of serving requests, cloud VM migration and placement to co-optimize for datacenter portfolio cost and environmental objectives \cite{5598305, 6877627, Bianchini2014GreenComputing, zhenhua2011Sigmetrics, zhenhua2015IEEENetw, zheng2020, 6578864, VirtualInterconnectorKelly2016, lawcarbonkubernetes}. Often, the power efficiency and environmental benefits are achieved by powering down redundant machines. Also, it is commonly assumed that datacenters are colocated with renewable assets, and the studied optimization models often incorporate forecasts of the onsite renewable supply. The implications of power rebalancing on energy markets, renewable portfolio, customer latency, and avoided carbon dioxide emissions were discussed in \cite{VirtualInterconnectorKelly2016}. In addition to real-time carbon-aware datacenter power management, \cite{Bianchini2014GreenComputing, 6298199} investigates long-term datacenter planning that cost-effectively “follows” renewables.  The first-at-scale system that uses real-time marginal carbon intensity at cloud clusters' grid locations to decide where to place cloud VMs was recently announced in \cite{carbonawarekubernetesblog}. 

The previous treatments of temporal shifting of flexible compute jobs were mainly addressed using an optimization-based framework with stylized models for job-level resource demand modeling, which typically include deadline constraints \cite{zhenhua2012Sigmetrics, 6114408, GreenHadoop}. Here, the goal is to delay jobs to greener hours of the day, where the delay decisions are the result of trade-offs between environmental, cost, and modeled performance objectives. These studies have mostly been theoretical in nature, with numerical demonstrations that use trace-driven simulations or small-scale prototypes with specific workloads.

Our work bears the closest resemblance to the approach in \cite{zhenhua2012Sigmetrics} in that the flexible workload is delayed using day-ahead capacity allocation that minimizes the next day's expected targeted cost and environmental objective. The key distinction between the approach in \cite{zhenhua2012Sigmetrics} and the one proposed in this paper lies in how the real-time workload management is designed to respond to capacity allocations. While the mechanism in \cite{zhenhua2012Sigmetrics} modifies time-based job scheduling decisions, Google's Carbon-Intelligent Computing is scheduler-agnostic in that its optimization outcome only changes the scheduler's perception of available resources, with no impact on the scheduling policy itself: the scheduler can decide what to queue, if needed. 

Rather than using stylized models of demand uncertainty and its translation to power consumption, Google's Carbon-Intelligent Computing system uses aggregate cluster-specific resource demand forecasts and power models trained separately for each cluster. Thus, the new framework captures diversity in workloads and hardware configurations at Google scale. 

One item that differentiates this new methodology from previous approaches is its inclusion of risk associated with application and infrastructure performance expectations. Special attention is given to (i) predicting the next day's flexible and inflexible compute usage, (ii) translating it to power consumption \cite{radovanovic2021power}, and (iii) then optimizing in a risk-aware manner. To meet Google's infrastructure and application SLOs, there is monitoring, performance tracking, and a feedback loop that evaluates the recent application level impact and controls load shaping accordingly.

\section{Datacenter architecture and cloud computing overview}\label{sec:overview}

Most of Google's compute resources reside in Google-designed datacenters with proprietary power distribution, cooling, networking and compute hardware \cite{45305, barroso2018datacenter}. The premise of carbon-aware computing based on day-ahead planning is that a known amount of computing, translated into power usage and optimized based on expected grid carbon intensity, yields the best placement of work. Furthermore,  managing peak power requires a good understanding of how workload resource usage translates to power. Therefore, it is important to have a good model of how resource usage and power inter-relate, and this requires a from-the-basics model of the complete datacenter power architecture. 

\subsection{Power architecture}\label{subsect:power_archit}

\begin{figure}
  \centering
  \includegraphics[width=\linewidth]{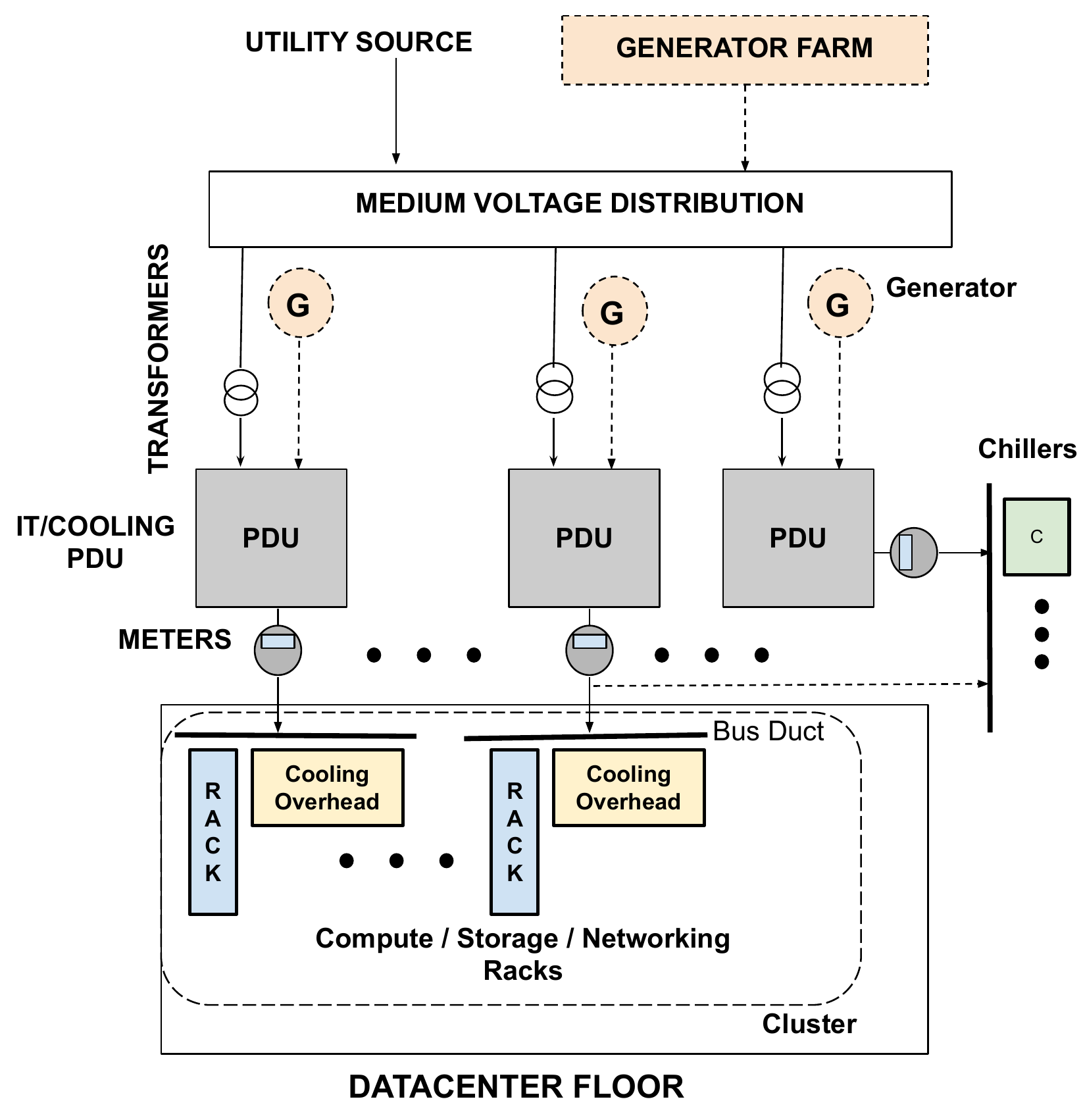}
  \caption{A simplified representation of a datacenter power distribution hierarchy.}
  \label{power_arch}
\end{figure}

Figure \ref{power_arch} shows a simplified view of power architecture of a typical Google datacenter \cite{barroso2018datacenter, radovanovic2021power}. Every datacenter is connected to the electricity grid via several medium voltage feeders. Each medium voltage distribution line is transformed to supply low voltage Power Distribution Units (PDUs). PDUs are connected to bus ducts. The bus ducts supply power to the IT and cooling equipment.

The IT equipment on the datacenter floor comprises compute, storage, and networking racks. A single Power Domain (PD) typically has a few thousand machines, and a handful of PDs comprise a cluster. Each PD is metered at a single PDU. The PDs in each cluster belong to a single job-scheduling domain, i.e., a common real-time scheduler that assigns computing tasks to its feasible machines. Generator backup is available to keep the datacenter running in the event of a grid power outage.

\subsection{Google's real-time resource management and its reliability principles} \label{sec:borgoperation}

Machines at Google are set up to run any application, and connected via high bandwidth switches within a campus, and via a global backbone of network connecting datacenters. Datacenter hardware is controlled and administered by specialized software that can handle massive scale. To the extent possible, hardware controls, job scheduling, etc., are abstracted away from users. 

Compute jobs at Google are managed by a distributed cluster-level operating system (known as Borg \cite{43438, 49065}). These jobs can be roughly split into two categories: (i) indefinitely running servers, and (ii) batch processes (e.g., data processing pipelines using MapReduce or Flume \cite{mapreducepaper, dataprocessingblog}). Jobs can consist of several tasks(sometimes thousands), both for reasons of reliability and because a single process can't usually handle all traffic. The cluster operating system is responsible for task allocation across machines within a cluster, which includes starting a job, finding machines for its tasks (i.e., task scheduling), allocating requested resources (CPU/RAM/disk) on machines, and instructing the machines to start executing the tasks. Since the scheduler needs to make hundreds of thousands of placement decisions per second \cite{49065}, it is important that the scheduling algorithm is computationally inexpensive and, therefore, fast, generally allowing jobs to flow into available compute resources like fluid into containers. The cluster operating system continually monitors and, potentially, reschedules tasks in case of problems. At Google, all available machines are typically turned on within a datacenter unless they are broken.

Every job belongs to a tier \cite{49065}, which is defined by its Service Level Objectives (SLOs). Higher tiers guarantee high availability of resources at all times with extremely small tolerance to disruptions, while lower tiers are designed to be used by applications that tolerate delays, e.g., creating new filter features on Google Photos, YouTube video processing, adding new words to Google Translate, and many other data processing and machine learning pipelines. Higher tier tasks have priority in real-time resource allocation/scheduling of machine resources within a cluster, and lower tier tasks are scheduled only if there are available resources remaining after scheduling the higher tier tasks. If resources are not available to run a job's tasks, they are queued. The admission controller visits this queue periodically, trying to enable jobs that pass a series of checks, one of which is cluster-level resource availability.

The resource requirements considered in scheduling a job's tasks are typically expressed using CPU, RAM, or disk units. At Google, CPU core requirements are expressed in Google Compute Units (GCU) \cite{49065, 9218509}, which hide differences across different hardware platforms. 1 GCU represents the amount of compute resources needed to generate the same amount of compute performance for a hypothetical “average” Google application across all hardware platforms. 1 GCU is easily translatable into an equivalent, fixed, number of platform specific CPU cores. For simplicity, for the rest of this paper we will refer to compute usage by the more generic term CPU.

In this paper, we divide tiers into 2 categories: temporally inflexible (higher tiers) and flexible (lower tier batch jobs that tolerate delays). We assume that the flexible workload assigned to a cluster can be subject to delays, as long as the amount of computation they perform during a day is preserved. Scheduling and the associated resource allocation is managed using the estimated upper bound of a task's actual usage across all resource dimensions (CPU, memory, disk, etc.). Our estimate of this upper bound, used for resource reservations, is greater than usage with high probability (close to 1). In prior work \cite{radovanovic2021power}, we demonstrated that the power usage of a cluster power domain can be estimated within 5\% error using a piecewise linear function of its CPU usage. As a result, any change in cluster-level CPU usage can be accurately mapped into a change in its power usage, which is critical for building CICS (see Subsection \ref{sec:powermodel} for more details).

Reliability considerations have played a major role in Google's system designs. Any software change or system configuration setting that can affect workload performance is subject to strict reliability guidelines. These prevent failures at a massive scale and ensure that, in cases where failures are detected, the changes can be pushed back and stopped in a timely manner. To that end, CICS's operation is monitored and load shaping is interrupted, in cases where the minimum flexible daily usage expectations are violated (more details are provided in Subsection \ref{sec:slos}).

The following subsection describes how the temporal flexibility of the flexible workload can be exploited to reduce carbon footprint and peak power usage of Google's datacenter portfolio. 

\subsection{Mechanism for carbon- and cost-aware load shaping}
\label{sbsec:mechanism}
This section discusses the framework we deployed for shaping intraday resource usage using time shifting of flexible workload in Google's compute clusters, with the goal to decrease both its electricity-based carbon footprint and improve overall resource efficiency (and, therefore, its long-term build costs).

The aggregate cluster-level load shaping is achieved using a Virtual Capacity Curve (VCC), which artificially limits the cluster's hourly compute usage and, in view of the power modeling advancements in \cite{radovanovic2021power}, its hourly power usage as well. Borg uses the VCC values to compute  the real-time CPU availability for incoming flexible jobs. Reducing cluster-level CPU capacity (i.e., total allowed CPU usage limits) at times of day when the corresponding grid carbon intensity is high or when it is cost-effective to do so, prevents the scheduler from starting as many jobs then, to reap the carbon or economic benefit. Flexible jobs get queued until resources become available. CPU load shaping affects resource usage across other datacenter resource dimensions as well (e.g., memory, disk, spindles, etc.). In aggregate, resource consumption is highly correlated to CPU consumption. Consequently, reduction of CPU consumption results in overall reduction of power consumption across all resources \cite{radovanovic2021power}. 

\begin{figure}
  \centering
  \includegraphics[width=\linewidth]{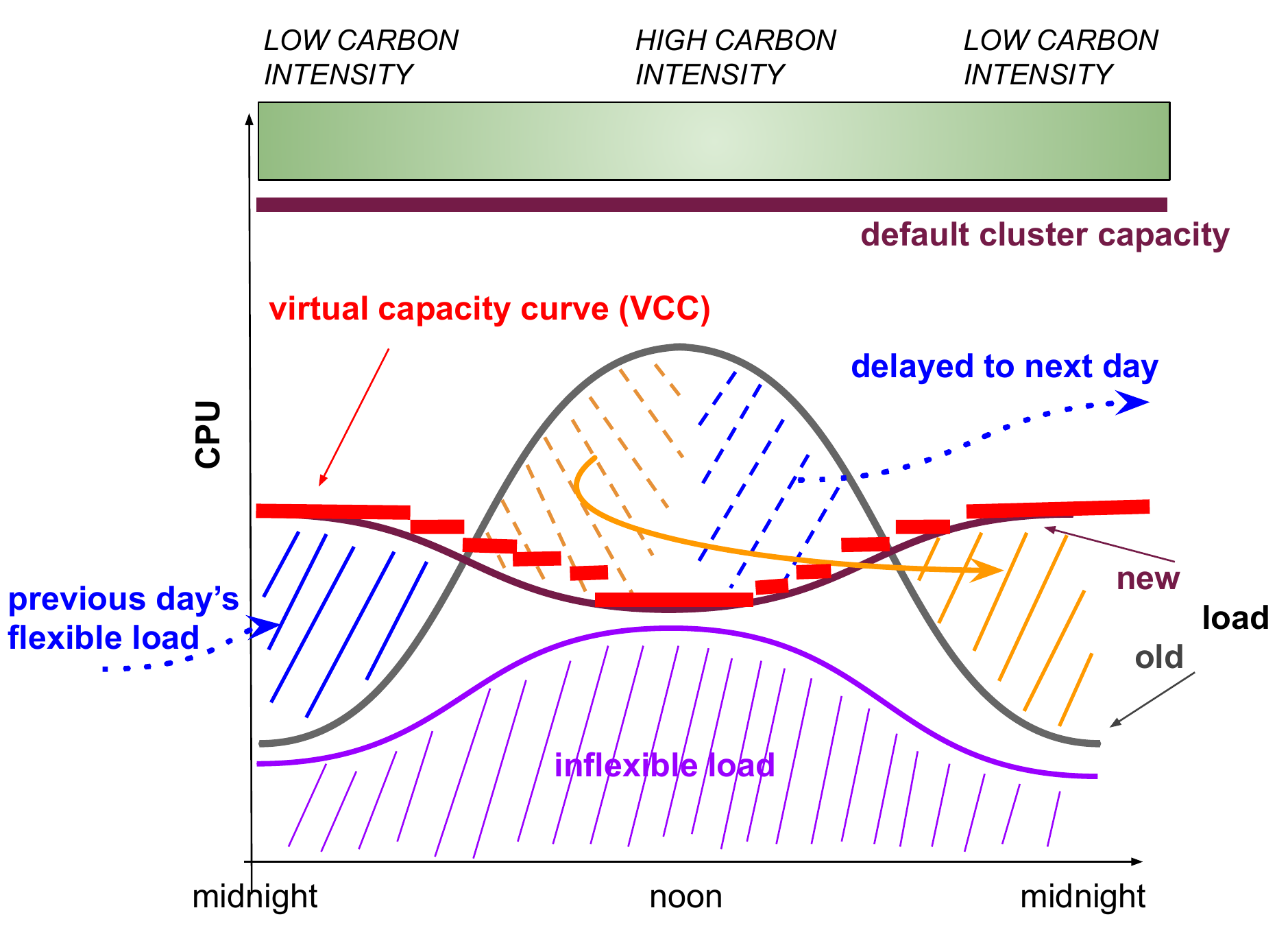}
  \caption{Effect of using the VCC mechanism for load shaping in a cluster.}
  \label{vcc_visual}
\end{figure}

An example of the effect that the VCC curve has on cluster CPU load shape is demonstrated in Figure \ref{vcc_visual}. VCC (in red) has lower values in the middle of the day when the intraday carbon intensity values are the highest. Flexible usage (orange and blue shaded areas) is pushed from midday to evenings and early mornings when the carbon intensity is lowest. In addition to carbon-aware load shifting in time, the proposed shaping mechanism  reduces daily peak CPU and, consequently, power consumption.

There are several key design principles that drove this system's design. 

\begin{itemize}
    \item \textbf{Limited Scope of Impact}: Optimal shifting should only impact flexible workload. Inflexible jobs should not be affected by time shifting.
    \item \textbf{User Impact Fairness}: When delaying execution of the flexible jobs, their users should be impacted in an unbiased way.
    \item \textbf{Reliability}: Changes to system configuration should be rolled out gradually so that the timely rollback is ensured in case of failures or disruptions to make sure that the fleetwide infrastructure and applications' SLOs are preserved.
    \item \textbf{Ramp-down Period}: The cluster operating system should know future VCC values in advance, so that it can stop allocating resources in a timely manner to achieve the desired CPU usage drops over time. 
    \item \textbf{Safety}: The time granularity of the VCC and, therefore, the smallest time between two consecutive computations and resubmissions of its values needs to allow for (i) sanity/safety checks (e.g., flexible load daily usage violation check, feasibility of VCC values, etc.) implemented within the cluster operating system, (ii) gradual rollout of VCCs to clusters across the fleet, and (iii) the explicit feedback loop if the safety checks are not satisfied. With that in mind and in view of the potential extensions to energy market applications, the VCCs are computed daily, and have hourly values that are used to shape the next day's load across all clusters fleetwide.
    \item \textbf{Modularity}: All data processing, forecasting and optimization pipelines used to compute day-ahead VCCs should be decoupled from the cluster operating system such that their failure would not impact the operation of the real-time cluster scheduler.
\end{itemize}

\section{Load shaping analytics}
\label{sec:shapinganalytics}

In this section, we describe the methodology for computing cluster-level VCCs. The implemented system comprises several components that: (i) predict the next day's load, (ii) train models that map CPU usage to power consumption, (iii) retrieve the next day's predictions for average carbon intensities on electrical grids where Google's datacenters reside, (iv) run day-ahead, risk-aware, optimization to compute VCCs, and (v) check for flexible workload SLO violations and trigger a feedback mechanism. As described earlier, the computed curves are used to reshape the corresponding cluster's intraday CPU usage and, therefore, power consumption (see more details below) by affecting only flexible (e.g., batch) workload. 

\begin{figure}
  \centering
  \includegraphics[width=\linewidth]{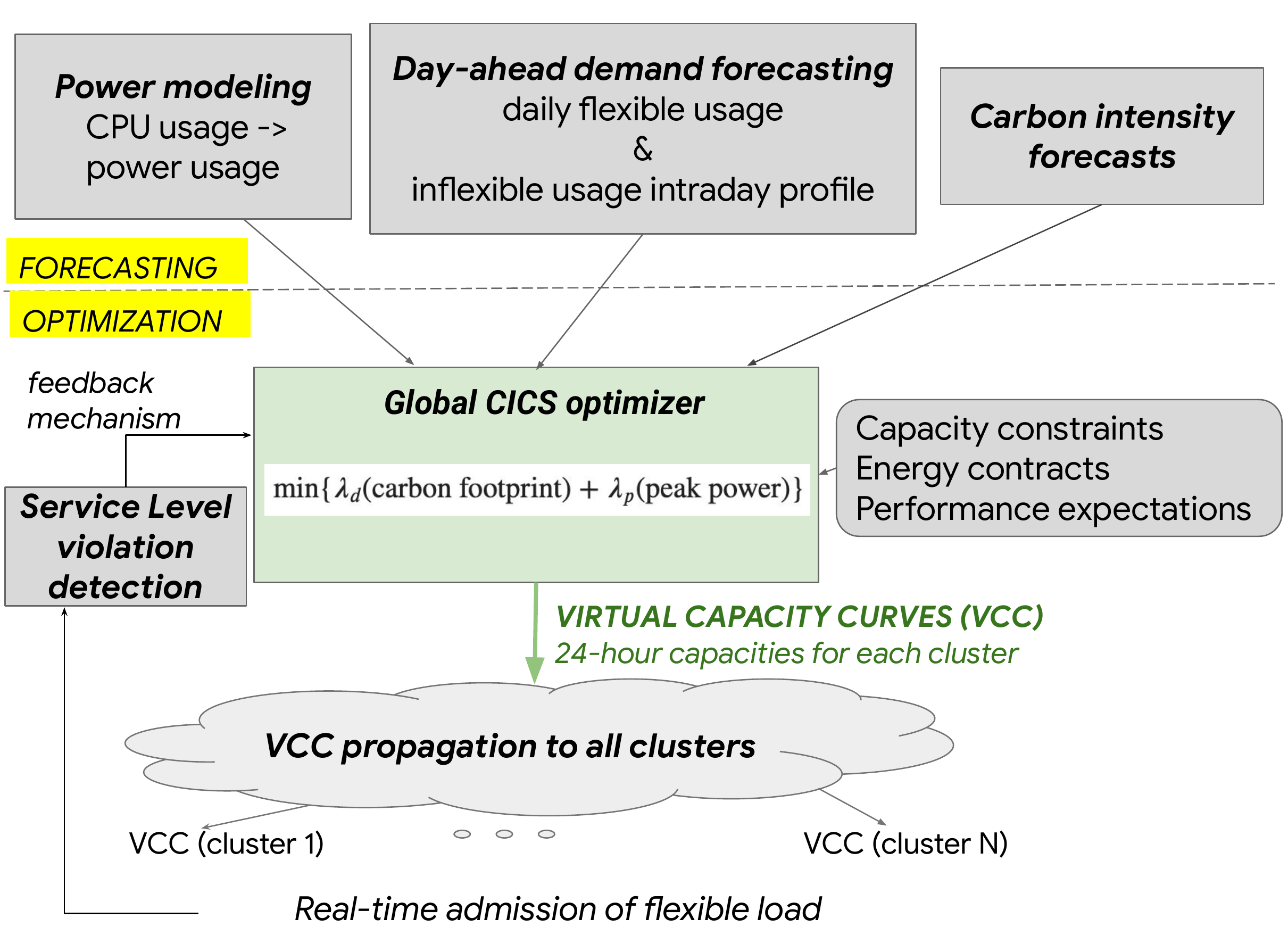}
  \caption{Architecture of the Carbon-Intelligent Computing system.}
  \label{achitecture}
\end{figure}

In view of the above, the suite of analytics pipelines (see Figure \ref{achitecture}) encompasses:

\begin{itemize}
    \item \textbf{Carbon fetching pipeline}, which reads hourly average carbon intensity forecasts from Tomorrow \cite{tomorrowwebsite} for each electricity grid zone where Google's datacenters reside. The forecasts are refreshed every hour. 
    \item \textbf{Power models pipeline}, which trains statistical models to map CPU usage to power consumption for each power domain (PD) across Google's datacenter fleet. The models are retrained and evaluated daily in a parallelized manner, and are demonstrated to be highly accurate irrespective of the underlying power architecture and machine platforms installed within a PD \cite{radovanovic2021power}. The ability to accurately translate changes in power domains and the corresponding clusters CPU usage to changes in power consumption enables carbon-aware load shaping, and is the core component of the Carbon-Intelligent system (for more details, see Section \ref{sec:powermodel} below).
    \item \textbf{Load forecasting pipeline}, which generates the next day's cluster-level forecasts for flexible and inflexible workload demand. Since inflexible load cannot be shaped, we predict its next day's hourly usage profile and the associated uncertainty. On the other hand, flexible load is considered shapeable as long as its total daily compute (CPU) demand is preserved. To that end, we predict the next day's flexible load compute usage, which turns out to be far more predictable than its typical daily usage profile. As discussed in Section \ref{sec:borgoperation}, there is a difference between the actual and reserved workload usage. Each cluster's real-time admission controller needs to ensure that every job's runtime reservations (i.e., actual usage upper bound) is met at any point of time throughout its execution. Thus, while the actual CPU usage directly drives power usage, the workload is required to be provided more capacity equal to its total reservations. The forecasting pipelines include accurate models of the relationship between the total cluster workload usage and reservations. It is particularly important that, alongside the expected values, the pipeline characterizes the forecasting error as well, which is embedded within the proposed risk-aware optimization framework to make sure that Google infrastructure and applications' SLOs are preserved. More on load forecasting methodology and its effectiveness is included in Section \ref{sec:loadforecasting}.
    \item \textbf{Optimization pipeline}, which runs daily to co-optimize the next day's fleetwide expected carbon footprint and appropriately-scaled power peaks, subject to infrastructure and application SLO constraints, contractual and resource capacity limits, as discussed in Section \ref{sec:optimization} below. The results of the optimization are cluster-level virtual capacity curves setting optimal capacity values for each hour of the next day. We use $VCC^{(c)}(h)$ to denote the optimal virtual capacity for cluster $c$ at hour $h$ of the next day. As previously discussed, new hourly capacities only affect flexible load, by queuing resource allocation requests and disabling some of the running tasks at hours when $VCC^{(c)}(h)$ values are low, which typically aligns with hours of day when carbon intensities are high, or when inflexible usage spikes.
    \item \textbf{SLO violation detection} flags when a cluster's daily flexible demand is not met. The violation can happen due to the unpredicted growth in flexible or inflexible usage (which, consequently, reduces resource available for flexible, lowest tier, workload). When the measured cluster flexible daily demand starts to persistently exceed the computed violation threshold (see Section \ref{sec:slos} for details), the Carbon-Intelligent system stops shaping these clusters for a week, to allow load forecasting models to adapt to changes in demand.
\end{itemize}

\begin{figure}
  \centering
  \includegraphics[width=\linewidth]{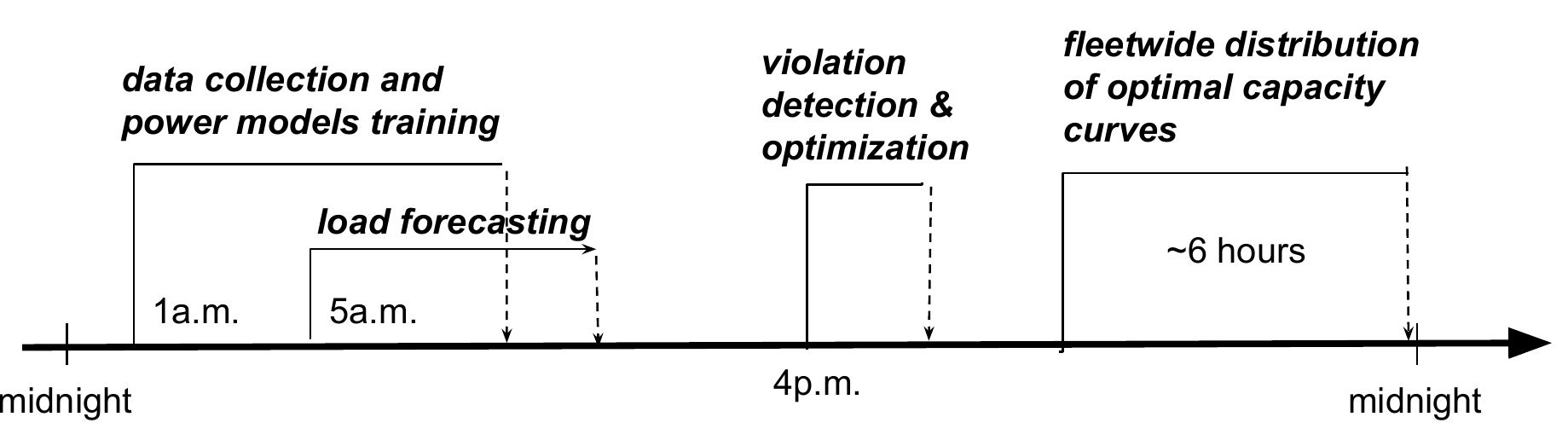}
  \caption{Timing of intraday analytics pipelines to ensure completion and reliable distribution of optimal capacity curves to all clusters across Google's datacenter fleet.}
  \label{pipeline_process}
\end{figure}

Daily analytics pipelines are scheduled at different times of a day. Data collection and modeling pipelines generate the next day's predictions (in particular, all usage data at Google is timestamped using Pacific Standard Time (PST)). These are then used by the central optimizer once per day to compute optimal next day VCCs for all clusters fleetwide. The generated VCCs span $24$ hours of the next day tracked in PST. An example of a schedule for pipeline processes is included in Figure \ref{pipeline_process}. 

In the rest of the section, we discuss the details of the implemented analytical approaches within the CICS. In the rest of the paper, we use capital letters to denote stochastic processes, and lowercase letters to denote their realizations. Also, the operator $\hat{\cdot}$ is used to denote the predicted value of a forecasted variable.

\subsection{Power modeling}
\label{sec:powermodel}

There have been many studies that suggest simple models to capture the relationship between machine/PD/cluster CPU usage and its power consumption \cite{fan2007power, radovanovic2021power, davis2011no, rivoire2008comparison}. The recent study \cite{radovanovic2021power}, evaluated across all of Google's power domains, demonstrated that a PD power consumption can be accurately estimated using only its GCU usage as a resource usage feature (which is a scaled version of the CPU usage). Moreover, this general conclusion was shown to hold irrespective of the hardware heterogeneity across power domains (e.g., diversity in machine types, such as compute, storage, accelerators, etc.)

More specifically, as discussed in \cite{radovanovic2021power}, a piecewise linear model accurately captures the relationship between CPU usage and dynamic power consumption for a given PD. Using a rigorous evaluation methodology and usage data at 5-minute time granularity, it was demonstrated that the daily Mean Absolute Percent Error (MAPE) of the proposed model is less than 5\% for more than 95\% of PDs. In our testing, we also find that the impact of a change in a PD CPU usage on power can be accurately locally approximated as
\begin{equation*}
\begin{aligned}
    Pow^{(PD)}(u_{CPU}^{(PD)} + \Delta u_{CPU}^{(PD)}) - Pow^{(PD)}(u_{CPU}^{(PD)}) \\ \approx \pi^{(PD)}(u_{CPU}^{(PD)})\Delta u_{CPU}^{(PD)}
\end{aligned}
\end{equation*}
where $Pow^{(PD)}(u_{CPU}^{(PD)})$ is used to denote PD power consumption when its CPU usage equals $u_{CPU}^{(PD)}$, $\pi^{(PD)}(u_{CPU}^{(PD)})$ is power model's slope at $u_{CPU}^{(PD)}$ for PD, and $\Delta u_{CPU}^{(PD)}$ denotes a change in CPU usage of power domain PD. More details on how the suggested piecewise linear model is retrained daily, how it adapts to changes in CPU and power usage regimes, and how its performance is evaluated can be found in III.A of \cite{radovanovic2021power}. Analogously, we use $u_{CPU}^{(c)}=\sum\limits_{PD \in c} u_{CPU}^{(PD)}$ to denote the total cluster $c$ CPU usage, and $U_{CPU}^{(PD)}(t)$ and $U_{CPU}^{(c)}(t)$ to denote PD and cluster CPU usage as a stochastic process in time, respectively.

The cluster-level workload scheduler assigns a large number of computing tasks in real time to randomly selected feasible machines in a given cluster. As a consequence, it is observed that the resulting CPU usage fractions across PDs within the same cluster $c$ varies insignificantly across time, i.e., 
$\frac{U_{CPU}^{(PD)}(t)}{\sum\limits_{PD \in c} U_{CPU}^{(PD)}(t)} \approx const$. Median variation, computed using the usage fractions for all PDs fleetwide collected at 5 minute time granularity, is estimated to be 1\%, where the larger variation values, close to 25\%, are recorded only for new clusters with very little workload. Therefore, by using $\lambda^{(PD)}$ to denote time average CPU usage fraction corresponding to the particular power domain, PD, i.e. $\lambda^{(PD)} = \mathbb{E} \left[ \frac{U_{CPU}^{(PD)}(t)}{\sum\limits_{PD \in c} U_{CPU}^{(PD)}(t)} \right]$, we obtain

\begin{equation}
\begin{aligned}
    & Pow^{(c)}(u_{CPU}^{(c)} + \Delta u_{CPU}^{(c)}) - Pow^{(c)}(u_{CPU}^{(c)}) \\
    & \approx \sum\limits_{PD \in c} \left (Pow^{(PD)}(u_{CPU}^{(PD)} + \Delta u_{CPU}^{(PD)}) - Pow^{(PD)}(u_{CPU}^{(PD)}) \right)\\
    & \approx \sum \limits_{PD \in c} \pi^{(PD)}(u_{CPU}^{(PD)}) \Delta u_{CPU}^{(PD)} \\
    & \approx \left( \sum \limits_{PD \in c} \pi^{(PD)}(u_{CPU}^{(PD)}) \lambda^{(PD)} \right) \Delta u_{CPU}^{(c)}.
\end{aligned}
\label{power_in_PDs}
\end{equation}

Therefore, we use \eqref{power_in_PDs} to define cluster $c$ power usage sensitivity with respect to its CPU usage as
\begin{equation*}
\begin{aligned}
    \pi^{(c)} (u_{CPU}^{(c)}) \equiv \sum \limits_{PD \in c} \pi^{(PD)}(u_{CPU}^{(PD)})\lambda^{(PD)}.
\end{aligned}
\end{equation*}

\subsection{Day-ahead forecasting}
\label{sec:dayaheadforecasting}
The proposed risk-aware optimization framework for computing VCCs requires a forward looking view of the next day's compute demand and carbon intensities. It also requires a method for detecting when flexible workload SLOs are not met, and mechanisms to respond to these events. The effectiveness of the proposed shaping in this paper is mainly due to the high prediction accuracy of the aggregated flexible and inflexible demands, as well as that of the next day's carbon intensities, provided by \cite{tomorrowwebsite}.

\begin{table}[]
\resizebox{\linewidth}{!}{%
\begin{tabular}{|p{0.25\linewidth}|p{0.75\linewidth}|}
\hline
 Variables  & Definition  \\ \hline
 $U_{IF}^{(c)}(h)$, $U_F^{(c)}(h)$  & Inflexible and flexible CPU usage of cluster $c$ at hour $h$, $h \in d$.  \\ \hline
 $R_{IF}^{(c)}(h)$, $R_F^{(c)}(h)$ & Inflexible and flexible CPU reservations of cluster $c$ at hour $h$, $h \in d$. \\ \hline
 $T_{U,F}^{(c)}(d)$ & Daily amount of flexible compute usage (in CPU-hour) of cluster $c$ on day $d$, defined as integral of cluster $c$ flexible CPU usage over day $d$. \\ \hline
 $T_R^{(c)}(d)$ &  Daily amount of CPU reservations of cluster $c$ on day $d$, defined as integral of all cluster $c$ reservations over day $d$.\\ \hline
 $\Theta^{(c)}(d)$ & cluster-level, SLO-based, capacity requirement, i.e. $\sum \limits_{h \in d} VCC^{(c)}(h)=\Theta^{(c)}(d)$.
  \\ \hline
 $\mathcal{R}^{(c)}(h)$ & CPU reservations-to-usage ratio ($\geq 1$) for cluster cat hour $h$, $h \in d$. It is a function of total cluster CPU usage. \\ \hline
 $\tau_U^{(c)}(d)$ & Risk-aware daily flexible compute usage for cluster $c$ on day $d$, $\tau_U^{(c)}(d) = \alpha^{(c)}(d) \hat{T}_U^{(c)}(d)$, where $\alpha^{(c)}(d)$ is a risk-aware inflation factor computed to ensure that Service Level Objectives are met (Subsection \ref{sec:slos}).
  \\ \hline
\end{tabular}%
}
\label{notation_table}
\caption{Notation.}
\end{table}

\subsubsection{Load forecasting}
\label{sec:loadforecasting}

\begin{figure}
  \centering
  \includegraphics[width=\linewidth]{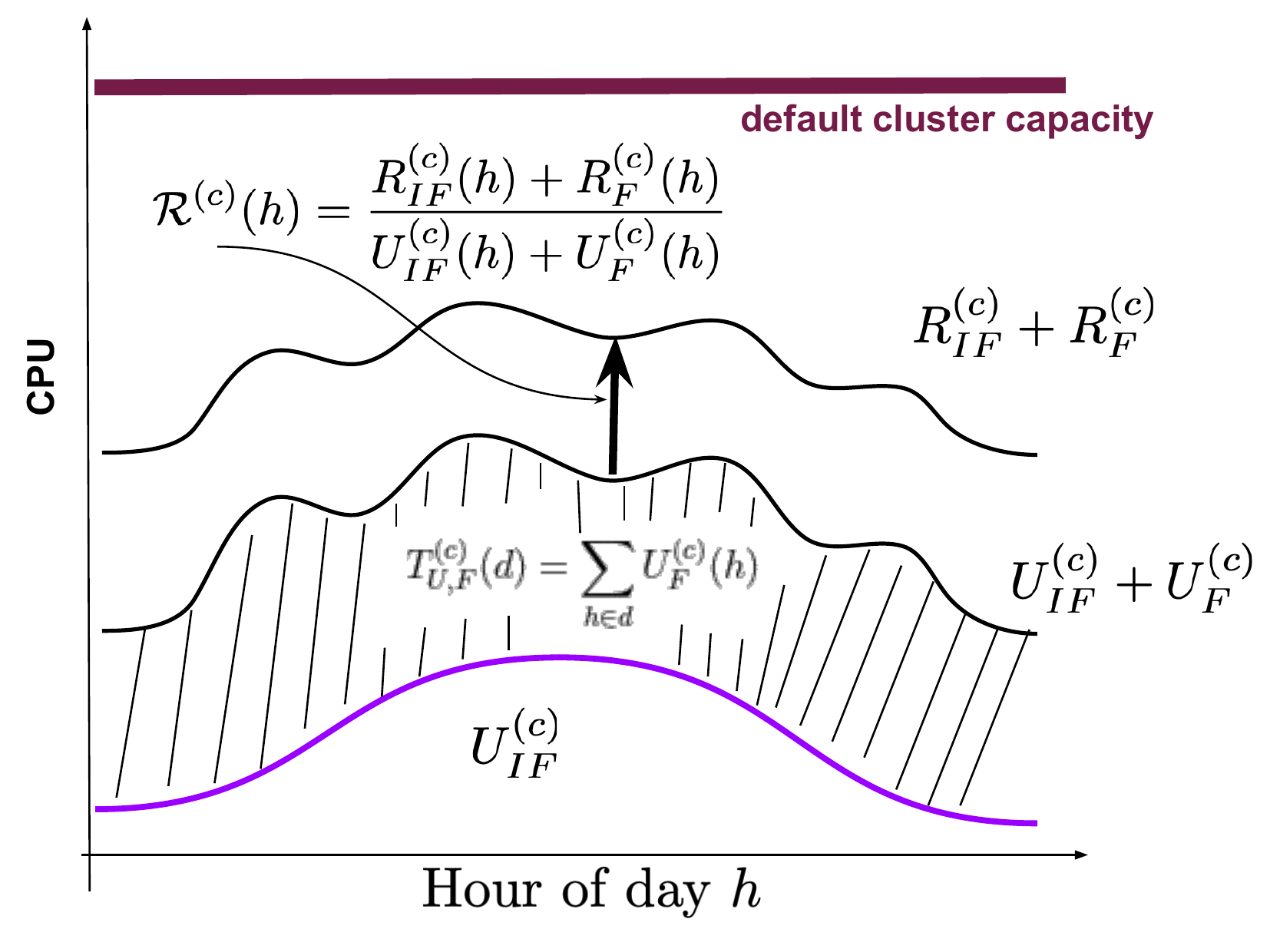}
  \caption{Predicted load components for each cluster $c$ fleetwide: inflexible usage daily profile $U_{IF}^{(c)}(h)$, daily flexible compute usage $T_{U,F}^{(c)}(d)$, all load reservation to usage ratio $\mathcal{R}^{(c)}(h)$.}
  \label{load_notation}
\end{figure}

The day-ahead forecasting pipeline of Google's Carbon-Intelligent Computing system predicts next day cluster-level: (i) hourly inflexible compute (CPU) usage, $U_{IF}^{(c)}(h)$, (ii) daily flexible compute usage, $T_{U,F}^{(c)}(d) =\sum \limits_{h \in d} U_F^{(c)}(h)$, (iii) daily total compute reservations, $T_R^{(c)}(d) = \sum \limits_{h \in d} (R_{IF}^{(c)}(h) + R_F^{(c)}(h))$, and (iv) ratio between total workload reservations and usage, denoted as $\mathcal{R}^{(c)}(h)$. The forecasted load components and their relationships are illustrated in Figure \ref{load_notation}.

The next day $d$ inflexible hourly CPU usage, $U_{IF}^{(c)}(h), h \in d$, daily flexible compute usage, $T_{U,F}^{(c)}(d)$, and daily total compute reservations, $T_R^{(c)}(d)$, are forecasted using a two-step approach. First, we predict their weekly average. Then, we augment these forecasts based on the deviation of the previous day's measurements from the corresponding weekly predictions.

Weekly forecasts for cluster-level, inflexible usage profiles, $U_{IF}^{(c)}(h)$, are obtained as a product of the next week's predictions for weekly mean value and hourly factors. Weekly mean value is forecasted using Exponential Weighted Moving Average (EWMA), with a half-life of $0.5$ (i.e., decay rate equal to $0.45$). Intra-week hourly factors are computed by dividing historical hourly usage by the corresponding weekly mean value. For each hour within a week, the hourly factors are forecasted using EWMA with a half-life of $4$ (i.e., decay rate equal to $0.07$). To compute weekly cluster-level forecasts for daily flexible compute usage and all-load compute reservations, i.e., $T_{U,F}^{(c)}(d)$ and $T_R^{(c)}(d)$, the analogous approach is used, with daily factor predictions instead of hourly. The EWMA parameters are selected by exploration over a given range, so that out-of-sample Mean Absolute Percent Error (MAPE) is minimized.

To adapt to intra-week deviations from the generated weekly forecasts, we observe that a simple linear model is able to capture the effect of the previous day's deviations from the weekly forecasts on next day's deviations. The predicted deviations are then added to the weekly forecasts for $U_{IF}^{(c)}(h)$, $T_{U,F}^{(c)}(d)$ and $T_R^{(c)}(d)$ to obtain final predictions used by the optimizer as discussed in Section \ref{sec:optimization} below.

It is observed that the ratio between the all load CPU reservations and usage, $\mathcal{R}^{(c)}(h)$, is primarily driven by the compute (CPU) usage. In particular, the larger the CPU usage of a cluster is, the smaller the ratio. To capture the observed trend, a linear model is trained which predicts reservation to usage ratio (greater than 1) as a function of log usage. The computed ratios are used by the optimizer to translate the computed next day optimal usage profiles into VCCs.

\begin{figure}
  \centering
  \includegraphics[width=\linewidth]{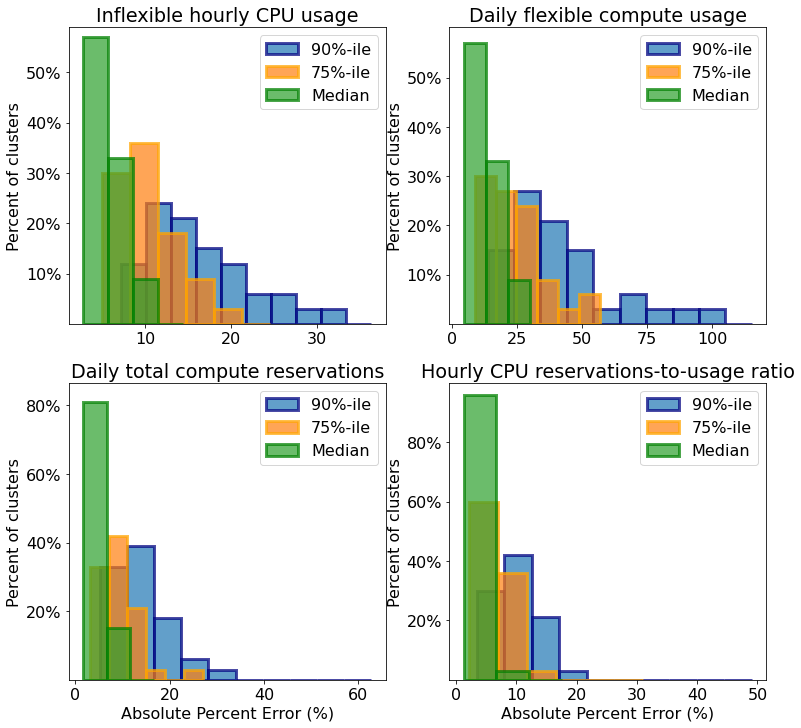}
  \caption{Percent of clusters (axis y) with median, 75\%-ile and 90\%-ile Average Percent Errors (rounded to the nearest 3\% increment) within a given range set by x axis. The x axis shows absolute percent error, and while outliers are in the 50-100\% range, the majority have far smaller errors. Consequently, these results suggest that the load is predictable}
  \label{forecast_results}
\end{figure}

The load forecasting models are trained and evaluated daily for all clusters across Google's datacenter fleet. For each cluster, we compute Absolute Percent Errors (APE) of all day-ahead predictions across a chosen 3-month-long time horizon. Then, we compute their median,  75\%-ile, and 90\%-ile, and we plot the distribution of their values for all clusters fleetwide. The results are shown in Figure \ref{forecast_results} for hourly inflexible CPU usage ($U_{IF}^{(c)}(h)$), daily flexible compute usage ($T_{U,F}^{(c)}(d)$), daily total compute reservations ($T_R^{(c)}(d)$), and hourly reservations-to-usage ratio ($\mathcal{R}^{(c)}(h)$) predictions. 

We can see that the median APEs of inflexible usage, total load reservations, and reservations-to-usage ratio predictions are smaller than 10\% for more than 90\% of the clusters. Daily flexible compute usage forecasts have larger APEs, which is not surprising given that flexible demand is typically more variable. The rare high (greater than $50$\%) APEs are sometimes observed for clusters that have small flexible usage, or that are going through a typically sudden, transient increase in flexible demand that happens due to infrastructure upgrades. These rare scenarios result in either inactive or ineffective shaping of the specific clusters on the particular days (see Subsection \ref{sec:optimization}), and are omitted when Figure \ref{forecast_results} was created.

\subsubsection{Service Level Objective awareness}
\label{sec:slos}

The main constraint within the implemented load shaping framework corresponds to workload SLO expectations, i.e., when a temporally flexible workload is shifted in time, its cluster-level daily compute usage must be preserved. The daily flexible compute usage is a stochastic process and, as previously discussed, is fairly predictable. In view of that, we set the SLO for its violation, which we explicitly embed into the proposed framework, and monitor for validation and safety. We define the target: the cluster-level daily amount of flexible compute (i.e., flexible load's daily capacity requirement) cannot be violated more often than one day per month when averaged across a given time horizon. This translates to roughly 3 days within a 100-day time horizon, which is equivalent to roughly 0.03 upper bound for violation probability. To meet this SLO, each cluster's VCC must ensure that the total amount of compute reservation demand satisfies

\begin{equation*}
\begin{aligned}
    \sum \limits_{h \in d} VCC^{(c)}(h) = (T_R^{(c)}(d))_{.97}
\end{aligned}
\end{equation*}

We compute the corresponding daily resource requirement, using the methodology discussed in Subsection \ref{sec:loadforecasting}. A violation of the introduced SLO is the result of two events: (i) the workload is granted more, reserved, CPUs than its actual usage to reliably run, and (ii) the total daily usage of temporally flexible, lower priority, workload is constrained either due to the unpredicted growth in its demand, or due to the unpredicted growth in the higher tier, inflexible workload.  

The 97\%-ile of the total daily capacity requirement (i.e., all-load daily compute reservations) is computed using the previous 90-day relative errors of the day-ahead predictions: 
\begin{equation*}
\begin{aligned}
    \Theta^{(c)}(d) & = (T_R^{(c)}(d))_{.97} \\ & = \left( \hat{T}_R^{(c)}(d) \right) \left( 1 + (\{\epsilon^{(c)}(n)\}_{n=d-90,...,d-1})_{.97} \right)
\end{aligned}
\end{equation*}
where $\epsilon^{(c)}(n)$ is the next day evaluation of the relative prediction error for day $n$, and $\hat{T}_R^{(c)}(d)$ is day $d$ prediction for next day's all-load compute reservations for cluster $c$. Thus, we ensure that

\begin{equation}
\begin{aligned}
    \sum \limits_{h \in d} VCC^{(c)}(h) = \Theta^{(c)}(d), 
\end{aligned}
\label{throughput_constraint}
\end{equation}
and, if the actual daily reservations demand gets close to the VCC limit for two days in a row, the system considers it a sign of cluster c daily usage violation, and triggers the feedback mechanism. While there are different ways to cope with the unpredicted demand growth, one option is to stop load shaping for some time (e.g., a week) to allow load forecasting models to adapt to changes in load demand.

To ensure that the CPU capacity limit in \eqref{throughput_constraint} is met in the optimal daily planning process (described in detail in Subsection \ref{sec:optimization} below), we attribute all the “extra” capacity to the daily amount of flexible usage by inflating its forecasted value with factor $\alpha^{(c)}(d)$ computed to satisfy
\begin{equation}
\begin{aligned}
    \sum \limits_{h \in d} \left( \hat{U}_{IF}^{(c)}(h) + \alpha^{(c)}(d) \frac{\hat{T}_{U,F}^{(c)}(d)}{24} \right) \hat{\mathcal{R}}^{(c)}(h) = \Theta^{(c)}(d)
\end{aligned}
\label{throughput_constraint2}
\end{equation}
where $\hat{\mathcal{R}}^{(c)}(h)$ is the predicted reservations-to-usage ratio corresponding to the nominal CPU usage at hour $h$, i.e. $\hat{U}_{IF}^{(c)}(h) + \frac{\hat{T}_{U,F}^{(c)}(d)}{24}$, on day $d$. In the rest of the paper, we use $\tau_U^{(c)}(d) = \alpha^{(c)}(d) \hat{T}_{U,F}^{(c)}(d)$ to denote the inflated, risk-aware, daily flexible usage.  

\subsubsection{Carbon intensity forecasting}
The optimization methodology embedded into our Carbon-Intelligent Computing system retrieves the near-term (48-hour) forecasts for average carbon intensities from Tomorrow (electricityMap.org). Tomorrow's approach accounts for demand, generation, and imports, to estimate the average carbon intensity of grid consumption in a particular region \cite{tomorrowwebsite}. To compute the optimal capacity plan for the next day, the optimizer uses the carbon intensity forecast for each location and hour of the next day, $\eta^{(c)}(h)$, obtained in kgCO$_{2}$e / kWh. The appropriateness of using the average carbon intensity vs any other indicator of the generation mix dispatch on the local grid (e.g., marginal carbon intensity) is discussed in more detail in Subsection \ref{sbsec:discussion}. Since a datacenter often contains many colocated clusters, the forecasted and actual carbon intensities are identical for clusters located in the same physical datacenter. The evaluated APE of Tomorrow's hourly forecasts used by the optimizer hugely depends on weather forecasts and the forecast horizon. Its mean value, MAPE computed for different grid locations where Google datacenters reside ranges between 0.4\% - 26\% over the range of forecast horizons (8-32 hours) for the next day forecast.

\subsection{Optimization framework}
\label{sec:optimization}
Our risk-, cost- and carbon-aware load shaping approach uses day-ahead forecasts to compute the next day's optimal capacity values (i.e., total reservation capacities) for each hour and cluster fleetwide. As discussed earlier, the optimal carbon-aware capacity limits are computed once daily and propagated gradually to adjust cluster configurations before the beginning of the next day. To that end, the uncertainty of the next day's predictions strongly impact the effectiveness of the proposed approach, and the proposed optimization methodology is carefully designed to harness predictable workload, environmental and infrastructure parameters. 

\begin{figure}
  \centering
  \includegraphics[width=\linewidth]{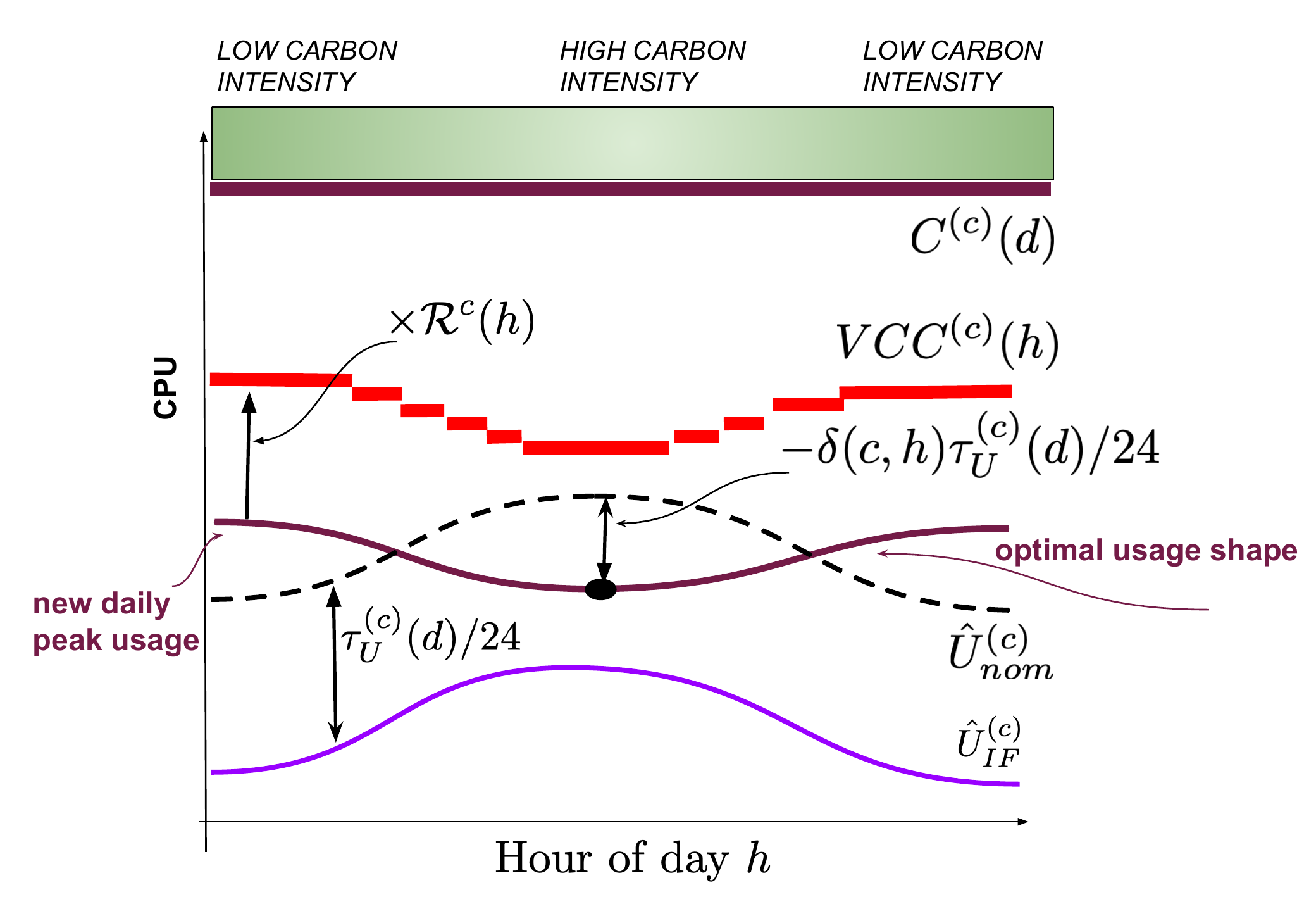}
  \caption{An example of using $\delta(c,\cdot)$ to control cluster-level CPU (and power usage) shape so that both its carbon footprint and daily usage peaks are reduced.}
  \label{optimiation_notation}
\end{figure}

The optimizer's objective is to derive next day's hourly reservation capacities that minimize the weighted sum of expected carbon footprint and daily power peak values summed over all clusters fleetwide, i.e.,

\begin{equation}
\begin{aligned}
    \min \limits_{\delta, y } & \text{ } \lambda_e \sum\limits_{c,h} \eta^{(c)}(h) ( Pow^{(c)}(\hat{U}_{nom}^{(c)}(h)) \\ & + \pi^{(c)} (\hat{U}_{nom}^{(c)}(h)) \delta(c,h)\frac{\tau_U^{(c)}(d)}{24})  \\ &  + \lambda_p \sum\limits_{c} y^{(c)}(d)
\end{aligned}
\label{objective}
\end{equation}

where $\lambda_e$ is the cost of 1 kg/CO$_2$e generated carbon footprint (\$ / kg CO$_2$e) and $\lambda_p$ is the cost associated with power infrastructure costs driven by clusters peak power consumption (\$ / MW / day). $\hat{U}_{nom}^{(c)}(h) \equiv \tau_U^{(c)}(d) / 24 + \hat{U}_{IF}^{(c)}(h)$ represents cluster $c$ nominal, risk-aware, CPU usage at hour $h$ of next day obtained by adding hourly prediction for inflexible CPU usage and the average hourly risk-aware flexible compute usage as defined in \eqref{throughput_constraint2}. Matrix $\delta$ (n x 24 matrix) is used to denote hourly deviations of flexible CPU usage from its average hourly target, $\tau_U^{(c)}(d)/24$; $y^{(c)}(d)$ is cluster $c$ upper bound for its daily peak power consumption. The goal is to compute optimal values for $\delta$ and $y^{(c)}(d)$. An example on how $\delta(c,\cdot)$ is used to control cluster-level CPU usage shape so that its carbon footprint and daily peak power usage are reduced, and how the optimal usage shape is translated into the corresponding VCC, is included in Figure \ref{optimiation_notation}.  The co-optimization ensures that the best possible outcome for total carbon footprint and infrastructure efficiency ensue. 

The optimization needs to ensure that application and infrastructure SLO constraints are honored, as well as contractual constraints that define the maximum datacenter power demand:

\textbf{The daily usage conservation constraint} for temporally flexible workloads translates into
\begin{equation*}
\begin{aligned}
    \sum\limits_{h} \delta(c,h) = 0, \forall c
\end{aligned}
\end{equation*}

\textbf{Power capping constraint} is a power infrastructure SLO constraint, which sets a threshold for cluster level CPU usage, $\bar{U}_{pow}^{(c)}$, to prevent power domains' circuit breakers from tripping (\cite{fan2007power, radovanovic2021power}). In particular, cluster $c$ compute usage can exceeded the given threshold with probability less than or equal to a given parameter $0<\gamma << 1$. Therefore, $\mathbb{P} \left[U_{IF}^{(c)}(h) + (1 + \delta(c,h)) \tau_U^{(c)}(d) / 24 \geq \bar{U}_{pow}^{(c)} \right] \leq \gamma$, or, equivalently,
\begin{equation*}
\begin{aligned}
    (U_{IF}^{(c)}(h))_{1-\gamma} \leq \bar{U}_{pow}^{(c)} - (1 + \delta(c,h)) \frac{\tau_U^{(c)}(d)}{24}, \forall c,h.
\end{aligned}
\end{equation*}
We use $(\cdot)_{1-\gamma}$ in the previous expression to denote $(1-\gamma)$th quantile evaluated using historical day-ahead predictions and actual, measured inflexible CPU usage, as discussed in Subsection \ref{sec:loadforecasting}.

\textbf{Campus-level energy contracts} set power usage limits for some Google datacenters, $L_{cont}^{(dc)}$, which the optimizer enforces by bounding the sum of cluster level peak power usage as
\begin{equation*}
\begin{aligned}
    \sum\limits_{c \in dc} y^{(c)} \leq L_{cont}^{(dc)}, \forall dc.
\end{aligned}
\end{equation*}

\textbf{Cluster-level total machine capacity}: the next day's optimal cluster-level CPU reservations profile cannot exceed its total machine capacity, $C^{(c)}(d)$. Therefore, in view of the previous discussion, the virtual capacity curve values are computed as
\begin{equation*}
\begin{aligned}
    VCC^{(c)}(h) = \left( \hat{U}_{IF}^{(c)}(h) + (1+\delta(c,h)) \frac{\tau_U^{(c)}(d)}{24} \right) \hat{\mathcal{R}}^{(c)}(h), \forall c,h,
\end{aligned}
\end{equation*}
where
\begin{equation*}
\begin{aligned}
    VCC^{(c)}(h) \leq C^{(c)}(d), \forall c,h.
\end{aligned}
\end{equation*}

\textbf{Other constraints}: note that there are other constraints that could be incorporated into the optimization. For example, a constraint could be added to bound the allowed drop in intraday flexible usage, or to bound hour-to-hour changes in $VCC^{(c)}(\cdot)$ values. Note that many of the listed constraints can be incorporated inside the objective terms (as soft constraints) using an appropriately large penalty and function form (e.g., hinge, quadratic, etc.). 

\subsection{Discussion}
\label{sbsec:discussion}

\textbf{Carbon vs peak power consumption cost:} The optimization outcome changes depending on the desired tradeoffs between carbon footprint and efficiency. Using only the carbon footprint as the objective term can potentially increase the machine capacity needed to meet the workload demand shifted to times of day when the local grid's carbon intensity is lowest. On the other hand, using a cost function that incorporates both carbon footprint and cluster-level power consumption peaks as in expression \eqref{objective}, Google decreases its load's expected global carbon footprint while also reducing demand for future infrastructure builds required to support its workload.

\textbf{Shaping using day-ahead average carbon intensities:} Day-ahead grid-level dispatch across different types of energy generation sources is fairly predictable and so are average carbon intensities \cite{8733097}. To that end, fairly consistent, planned reshaping of next day's datacenter power consumption that drops compute and, therefore, power usage at hours of a day when the average carbon intensity is predicted to be high, is expected to affect day-ahead dispatch decisions and result in lower overall CO$_2$ emissions. Note, however, that the choice of the most effective grid-level signal requires further research and data-driven investigation.    

\textbf{Comparison to Model Predictive Control (MPC):} MPC is a common approach used for ahead-of-time planning of load and generation dispatch in power systems \cite{DELREAL201465, 5759140, 7419888}. Hourly or more frequent dispatch updates of new operational set points are computed using regenerated demand and supply forecasts. MPC's robustness to the uncertainty in demand and generation is achieved via recurring updates. By contrast, the predictability of day-ahead flexible and inflexible compute and power usage is what makes the load shaping mechanism of CICS robust. In the unlikely case there is unpredictable growth in compute usage, CISC has an explicit feedback loop that disables cluster shaping for a week in case its daily flexible compute demand is not met.

\textbf{What makes the proposed mechanism effective?:} For Google's application, using cluster-level VCCs as a load shaping mechanism has several advantages over the previous academic proposals \cite{5598305, 6877627, Bianchini2014GreenComputing, zhenhua2011Sigmetrics, zhenhua2015IEEENetw, zhenhua2012Sigmetrics, 6114408, GreenHadoop}:
\begin{itemize}
\item \textit{Day-ahead vs real-time operations}: computation of day-ahead VCC curves is decoupled from the real-time scheduling operations, which need to be lightweight.
\item \textit{Computation depends on predictable optimization parameters}: VCCs are computed using a forward-looking view of CPU demand, carbon intensity forecasts and capacity limits, and are not affected by the intrinsic and significant uncertainty in flexible-job arrival patterns and resource usage. Computation of VCCs relies on significantly more predictable, aggregate flexible load demand and flexible workload daily usage.
\item \textit{Risk-awareness and extensibility}: the daily flexible usage conservation constraint for each cluster's flexible workload is explicitly defined. The CICS includes a mechanism for tracking cluster-level amounts of compute, detecting violations and responding to them. In addition to the cluster-level power capping constraints, and the explicit contractual bounds for datacenter power limits, the optimization formulation can be extended to include other hardware and workload properties and requirements (e.g., spatial shifting).  
\item \textit{Scalability}: this treatment using aggregate flexible and inflexible usage for centralized fleetwide optimization is significantly more scalable and computationally efficient than using typically diverse and hard-to-characterize job-level models. It  allows for future, spatial shifting, extensions, and portfolio-level resource optimizations. All this is achieved while meeting the reliability criteria, which is the most important design principle. 
\end{itemize}

\section{Demonstration and impact}
\label{sec:demo}

The impact of the carbon-aware computing approach can be observed across Google's entire fleet, spanning different electricity grids. The magnitude of these benefits, however, varies significantly from location to location. This section evaluates the impact of the proposed shaping mechanism by analysing operational data, showing how it is affected by (i) the amount of flexible usage, (ii) the high uncertainty range in the computed demand forecast, (iii) the intraday variability and magnitude of grid carbon intensity.

The following figures highlight how the predictability of compute demand impacts the effectiveness of load shaping. The carbon-aware computing mechanism aims to constrain load to limit power peaks and to shift compute away from hours with high carbon-intensity, while ensuring with high probability that all temporally flexible compute jobs will run within 24 hours.  

\begin{figure}
  \centering
  \includegraphics[width=\linewidth]{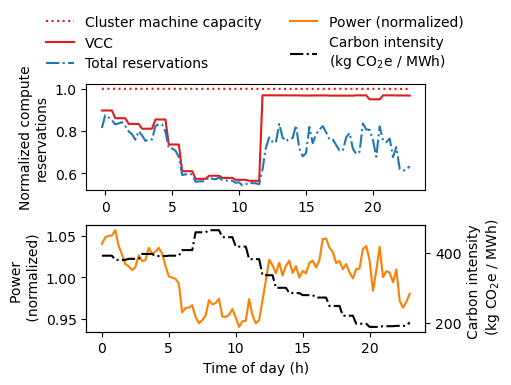}
  \caption{Hourly VCC, compute reservations, cluster power, and carbon intensity in cluster $X$ on the selected day.}
  \label{clusterX}
\end{figure}

\begin{figure}
  \centering
  \includegraphics[width=\linewidth]{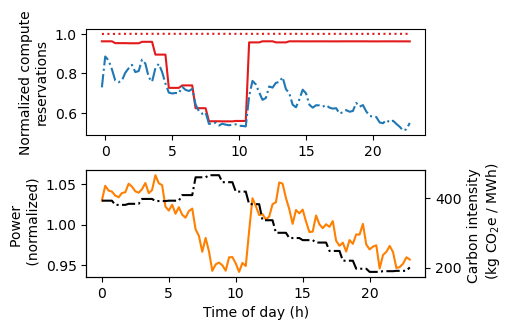}
  \caption{Hourly VCC, compute reservations, cluster power, and carbon intensity in cluster $Y$ on the selected day.}
  \label{clusterY}
\end{figure}

\begin{figure}
  \centering
  \includegraphics[width=\linewidth]{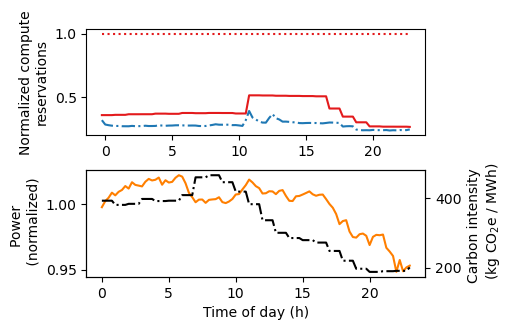}
  \caption{Hourly VCC, compute reservations, cluster power, and carbon intensity in cluster $Z$ on the selected day.}
  \label{clusterZ}
\end{figure}

This example depicts three clusters within a large Google campus on a selected day, included in Figures \ref{clusterX}, \ref{clusterY}, and \ref{clusterZ}. In each Figure, the top graph depicts real-time compute reservations (blue) constrained by a VCC (red). The bottom graph shows the normalized power used by the data center (orange), with the power's carbon-intensity (black) which influenced the VCC. The result: electricity usage was decreased precisely when the grid was the most carbon intense. 

In the first cluster, $X$, the average value of the VCC is about 18\% higher than the average load demand. This difference is due to the uncertainty in the load forecast, because our forecast value for the 97th quantile for load demand is 18\% higher than the actual daily load.

In this location, the VCC is able to drop flexible load by roughly 50\% during peak carbon intensity hours driving an 8\% drop in power during the hours with peak carbon intensity 
(see Figure \ref{clusterX}). In another location (see Figure \ref{clusterY}), $Y$, uncertainty in the forecasts for inflexible and flexible load drives the VCC higher. Here, the average value of the VCC is about 33\% higher than the average load demand. The VCC is still able to drop by almost 50\% during peak carbon intensity hours, but the drop is not as sustained (its duration is shorter). This drives a roughly 8\% decrease in power during the hours with peak carbon intensity, but the duration is only 3 hours versus 6 hours in cluster $X$, reducing the carbon impact of shaping. The higher predictability of load in cluster $X$ allows for more effective shaping and higher carbon reductions in cluster $X$ than in cluster $Y$. 

Next, the third example considers cluster, $Z$, where flexible compute load is small compared to inflexible load, captured in Figure \ref{clusterZ}. In this cluster, the extent of flexible load is relatively small compared to the cluster capacity and the extent of inflexible load. The latter relation implies that uncertainty regarding inflexible load demand, which shifts the VCC upwards, can prevent the optimization result from meaningfully shifting flexible load. In this cluster, there is no meaningful curtailment of flexible workloads on this day, and no significant change in power consumption during peak carbon intensity hours. 

Note that a VCC is set to a cluster total machine capacity when a cluster is too full to allow for shaping, for example when the risk-aware total daily compute reservations is larger than cluster machine capacity. The same happens when there is insufficient data for forecasting or estimating power models. These two scenarios typically happen in about 10\% of clusters on any given day. 

It is helpful to evaluate aggregate impacts on campuses and Google's fleet, due to differences in impact of the VCCs across datacenter locations. To that end, we ran a controlled experiment to evaluate the impact of day-to-day shaping on power consumption. Figure 13 shows the normalized power curves, averaged across all datacenter clusters in a campus, on randomly treated (optimized) and non-treated (not optimized) days for two months beginning February 12th, 2021. On each day, each cluster is randomly assigned to receive the carbon-aware optimal shaping or not, with 50\% probability of being in each group on any given day. 

\begin{figure}
  \centering
  \includegraphics[width=\linewidth]{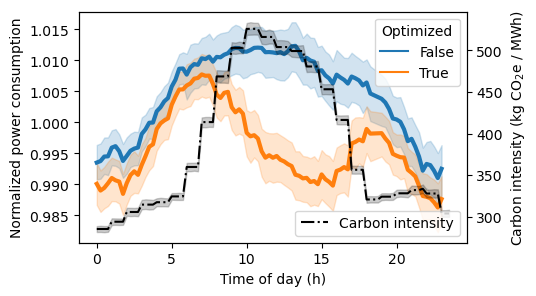}
  \caption{The effect of load shaping optimization on cluster power consumption, averaged across all clusters and test days in a single Google datacenter campus. The solid lines plot normalized power, over the course of the day, averaged over all data center clusters within a selected campus for 1 month, for shaped (orange) and not shaped (blue) clusters. The (black) dashed line displays the average carbon intensity in each hour over the course of the day for the grid that it is powered by. In each line, the uncertainty band reflects the 95\% confidence interval in the mean value for the hour, averaged across days and clusters.}
  \label{campus_impact}
\end{figure}

In this example, when CICS is active, average cluster power drops by 1-2\% during the highest carbon hours, compared to power in the same clusters on dates that the cluster is placed in the control group and not shaped by the carbon-aware computing mechanism. We also observe that, in load shaping regimes where we allow for \textit{larger and longer drops} in capacity (obtained by increasing the cost associated with the carbon footprint, $\lambda_e$, in \eqref{objective} and by relaxing conditions for $\delta$), total daily flexible compute usage tends to slightly decrease in the shaped clusters. Therefore, the total daily energy consumption tends to decrease when a cluster is shaped. This could be a consequence of the longer delay intolerance of some flexible workloads, and because some flexible jobs ``choose'' to move to different clusters in response to lower and durable virtual capacity limits imposed by the carbon-aware computing mechanism. As a result, in these ``more aggressive'' operating regimes, the daily flexible compute conservation condition fails to hold. The ``spontaneous'' load shifting to other locations may increase or decrease carbon emissions. To harness the spatial flexibility to carbon- and cost-effectively redistribute flexible load across both time and locations, future models will explicitly characterize spatially flexible demand and extend the proposed optimization framework to take it into consideration.

While one could set out to calculate the carbon impact using the power and carbon intensity data, that does not necessarily yield the correct result. There are different ways the impact can be evaluated and this will be investigated in future research. 

\section{Concluding remarks}
\label{sec:conclusion}

The growing climate impact of increased Greenhouse Gas Emissions and CO$_{2}$ levels in Earth's atmosphere highlights the value and importance of technologies that reduce such impact. Electricity generation is one of the larger contributors to global CO$_{2}$ emissions \cite{projectDrawdown}. The datacenter industry accounts for an expanding electricity demand, expected to reach anywhere from 3 to 13\% of global electricity demand by 2030 \cite{RePEc:gam:jchals:v:6:y:2015:i:1:p:117-157:d:49008}. Yet, it has the potential to facilitate grid decarbonization in a manner different from isolated power loads. There is an opportunity to harness datacenter load flexibility, not only to reduce emissions, but also to contribute to more robust, resilient, and cost-efficient decarbonization of the grid through energy system integration. 

This paper introduces Google's Carbon-Intelligent Computing System, which shifts datacenter computing in time and will soon also shift computing in space. These together will help realize the company's global environmental \cite{googlegoal} and efficiency objectives. The system proactively makes automated adjustments based on current and forecasted grid conditions to reliably and effectively shape Google's compute load in a carbon- and efficiency-aware manner. The core of the carbon-aware load shaping mechanism are cluster-level \cite{43438} Virtual Capacity Curves, which are hourly resource usage limits that serve to shape the cluster resource and power usage profile over the following day. These limits are computed using an optimization process that takes account of aggregate flexible and inflexible demand predictions and their uncertainty, hourly carbon intensity forecasts \cite{tomorrowwebsite}, explicit characterization of business and environmental targets, infrastructure and workload performance expectations, and usage limits set by energy providers for different datacenters across Google's fleet.

Using actual measurements from Google datacenter clusters, we demonstrate a power consumption drop of 1-2\% at times with the highest carbon intensity. Ongoing system enhancements, which include shifting flexible workloads across locations, are expected to increase the benefits of this system. Furthermore, due to this system's incorporated power models and the way virtual capacity curves are determined, future extensions could enable energy management within datacenter microgrids and integration with grid-level demand response programs.

The framework and principles embedded in Google's Carbon Intelligent Computing system align with its compute management systems and workload properties.  While other compute providers' approaches to carbon-aware computing will necessarily vary, we hope that the initial results demonstrated in this paper inspire academia and industry to pursue diverse approaches to individual cluster or hyperscale computing management.



\bibliographystyle{IEEEtran}
\bibliography{IEEEabrv,references.bib}
%

%








\end{document}